\documentclass[11pt,preprint]{aastex}
\newcommand{\pref}{\protect\ref} \newcommand{\solrad}{\ifmmode{R}_{\rm
S}\else${R}_{\rm S}$\fi} \newcommand{\solmas}{\ifmmode{M}_{\rm
S}\else${M}_{\rm S}$\fi}
\newcommand{\tintu}{$\,$ergs$\,$cm$^{-2}\,$s$^{-1}\,$sr$^{-1}$}

\newcommand{\ctn}{\ifmmode\kappa\else$\kappa$\fi}

\newcommand{\velu}{$\,$km$\,$s$^{-1}$}

\newcommand{\term}[2]{\mbox{$\,^{#1}{\rm #2}$}}
\def\term#1 #2/{\mbox{$\,^{#1}{\rm #2}$}}

\newcommand{\pop}{n}

\newcommand{\ddt}[1]{{ \partial~ \over  \partial t}#1 +  { { \partial~\over  \partial x}{v#1} }}
\newcommand{\frt}{(x,t)}
\renewcommand{\frt}{}

\slugcomment{}

\newcommand\tabone{
\begin{deluxetable}{lllllll}
\renewcommand{\baselinestretch}{1.0}
\tablecaption{\label{tab:parameters}
Parameters of the acceleration and heating }
\tablehead{Calculation & $t_1$  & $t_2$ & $z_1$ & $z_2$ & $h_1$ & $h_2$ \\
              & s & s & km  & km  & &}
\startdata
A & 30 & 30 & 2000 & 2400 & 3 & 30 \\
B & 15 & 15 & 1900 & 2500 & 3 & 100 \\
\enddata 
\tablecomments{$t_1$ is the duration of the acceleration and moderate
  heating phase, $t_2$
  the coasting and large heating phase.  The initial atmospheric plasma
between heights $z_1$ and $z_2$ is the region subject to the
acceleration and heating. $h_1$ and $h_2$ measure the factors by which
the nominal heating rates (equation \pref{eq:1}) are increased during $t_1$
and $t_2$ respectively. }
\end{deluxetable}
}

\newcommand\figone{
\begin{figure}[!ht] 
\epsscale{0.8}
\plotone{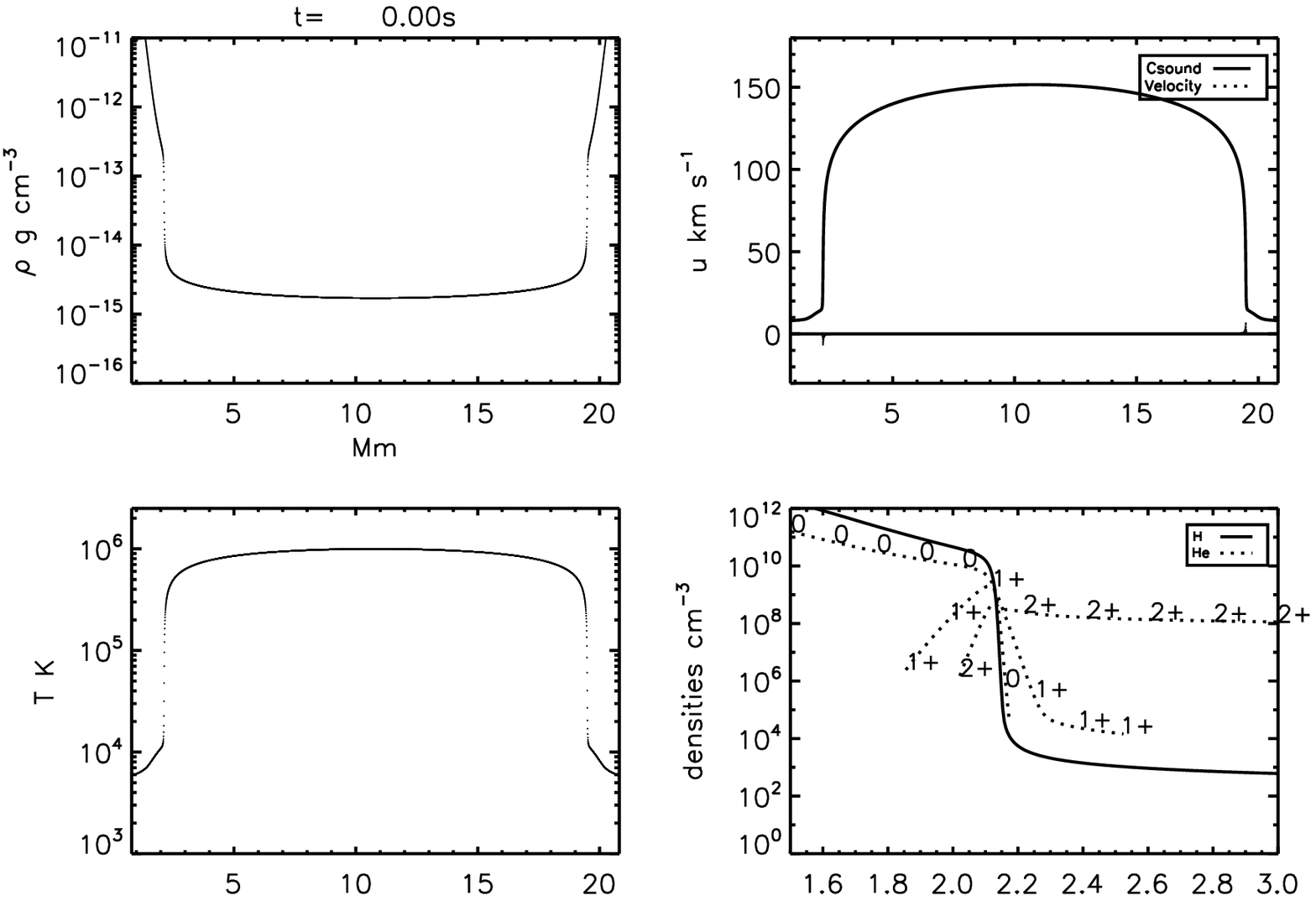}  
\caption{\label{fig:eqm} The initial, near-equilibrium state of the
  atmosphere is shown.  
}
\end{figure}
}

\newcommand\figtwo{
\begin{figure}[!ht] 
\epsscale{0.8}
\plotone{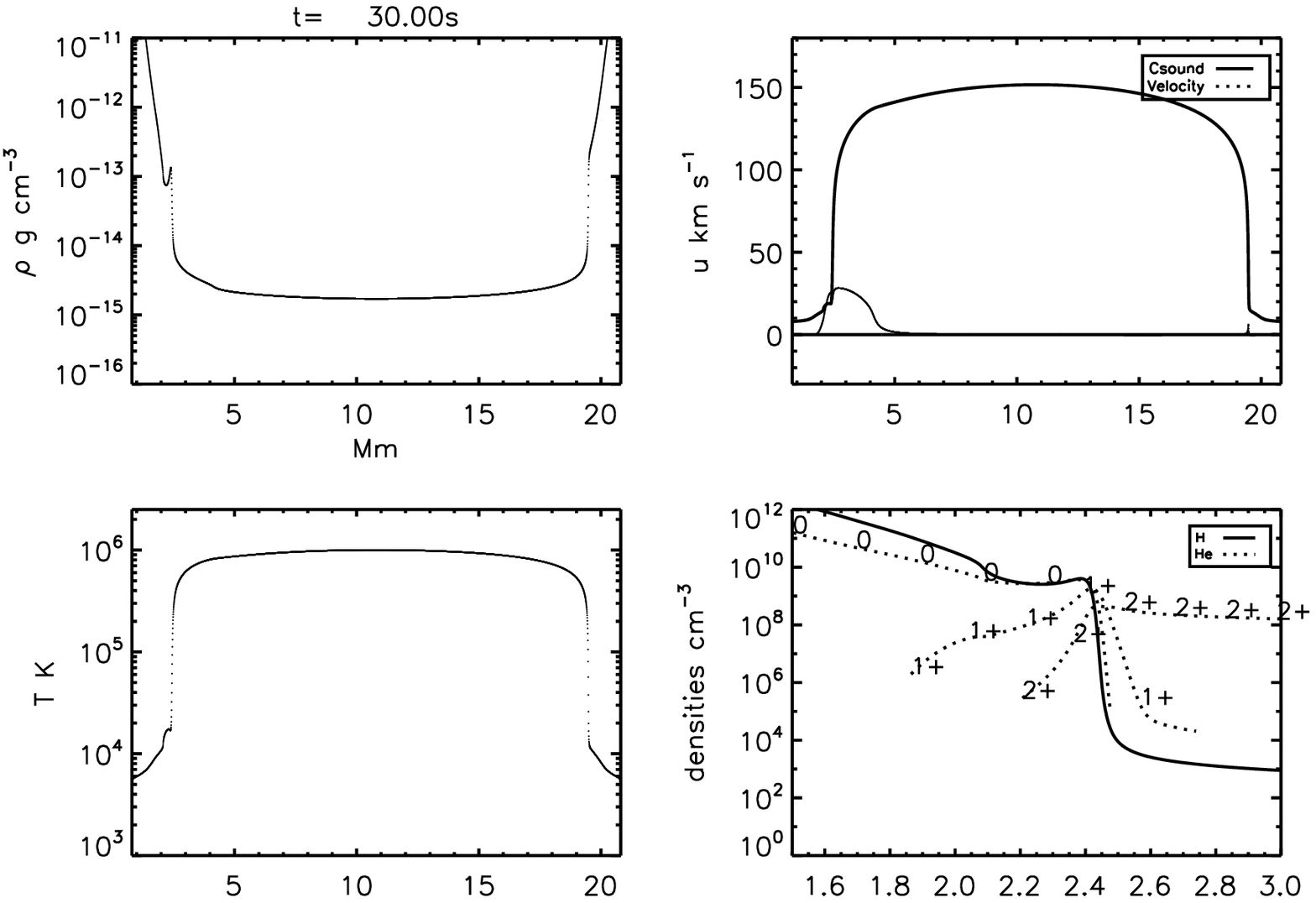}  
\caption{\label{fig:end} The state of the
  atmosphere after 20 seconds of acceleration and moderate extra
  heating in the upper chromosphere is shown, for calculation A.
}
\end{figure}
}

\newcommand\figthree{
\begin{figure}[!ht] 
\epsscale{0.8}
\plotone{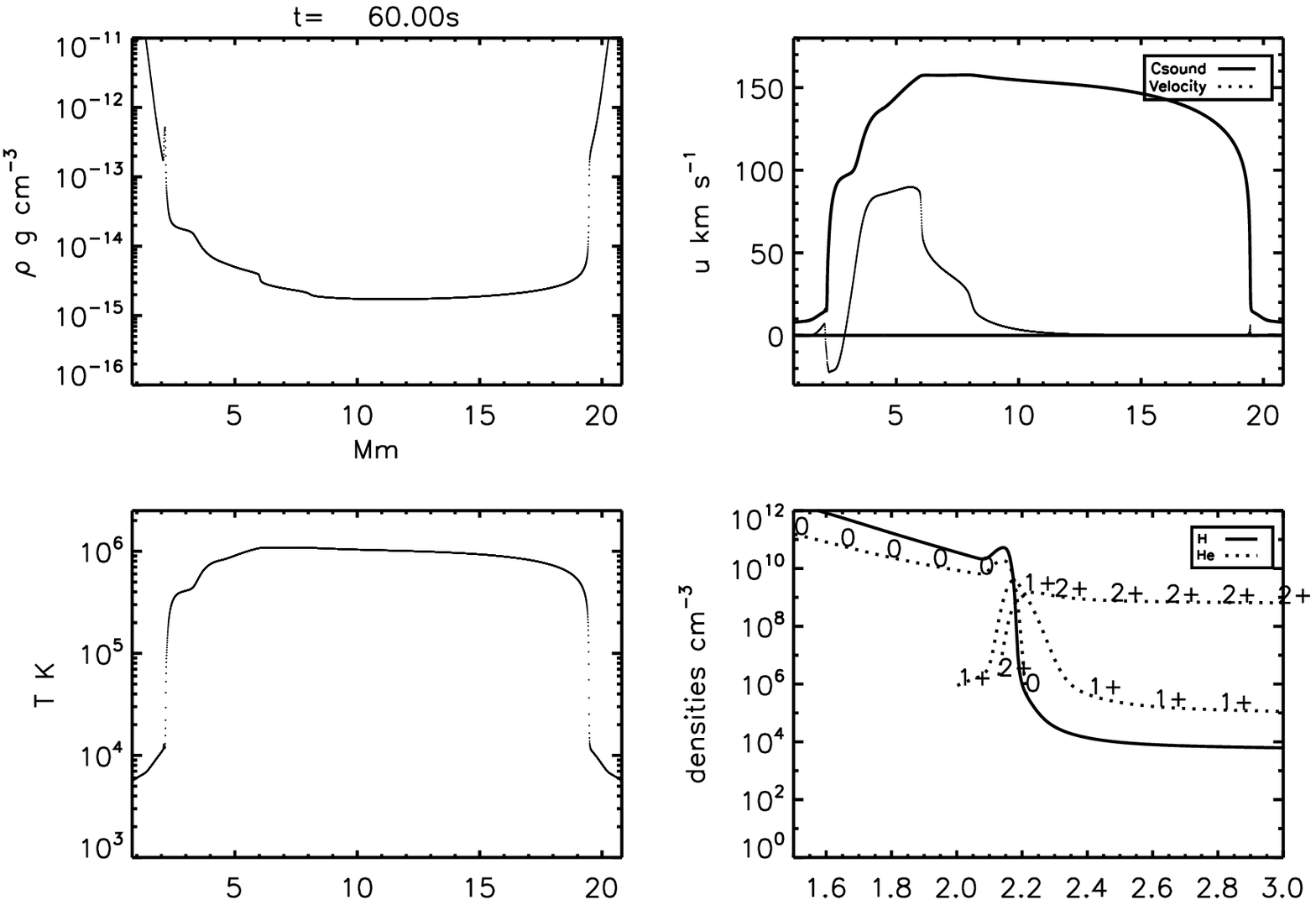}  
\caption{\label{fig:afterend} The state of the
 atmosphere 20 seconds into the intense heating phase is shown, for calculation A.
}
\end{figure}
}

\newcommand\figfour{
\begin{figure}[!ht] 
\epsscale{1.1}
\plotone{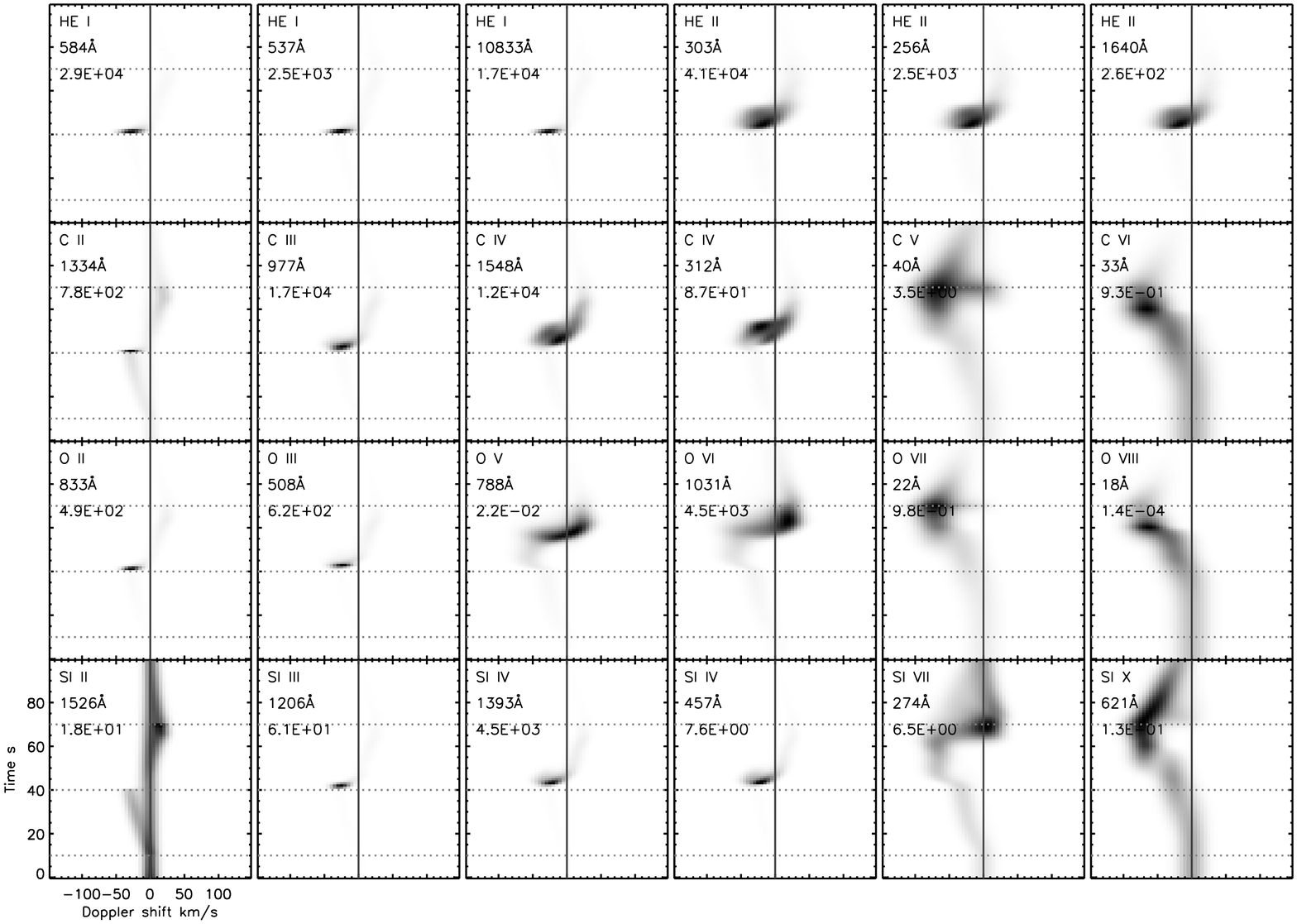}  
\caption{\label{fig:profiles} Spatially-averaged emission profiles
  from near the accelerated and heated footpoint of the loop, for
  calculation A. The emission was integrated along the vertical and
  averaged over all spatial positions within 2.175 Mm (corresponding
  to a 3 \arcsec{} spatial bin) of the footpoint in the low
  chromosphere. The panels are labeled with the atomic ion, central
  wavelength, and computed maximum total intensity (\tintu{}). 
  The horizontal dotted lines
  demark the periods of acceleration and heating (10 to 40s), and of
  more intense heating (40 to 70s).  }
\end{figure}
}

\newcommand\figfive{
\begin{figure}[!ht] 
\epsscale{1.1}
\plotone{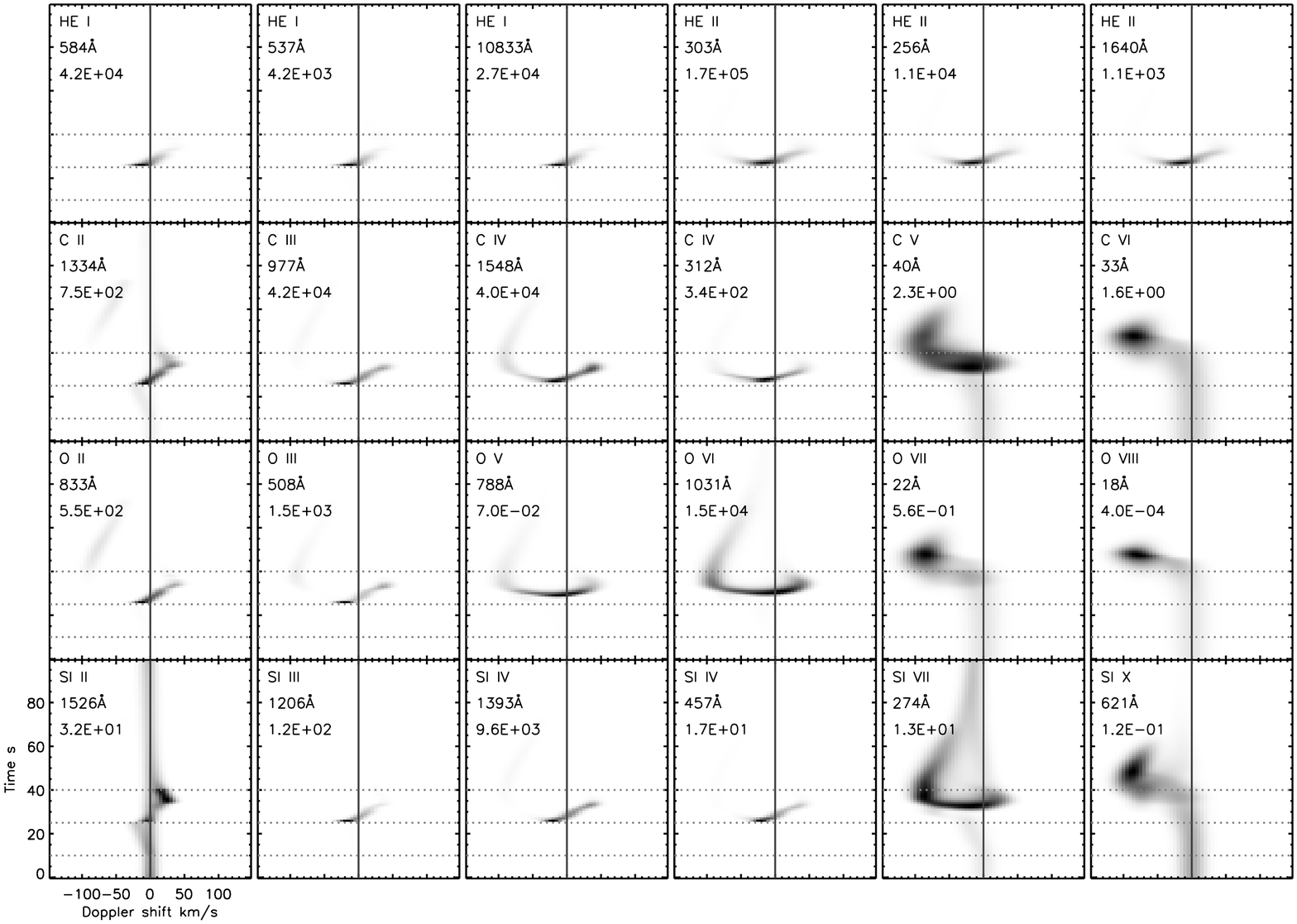}  
\caption{\label{fig:profiles1} Same as Figure~\pref{fig:profiles}, but
    for calculation B, which has higher heating rates in denser plasma
    than calculation
  A, and shorter periods of heating and acceleration.}
\end{figure}
}

\newcommand\figsix{
\begin{figure}[!ht] 
\epsscale{1.1}
\plotone{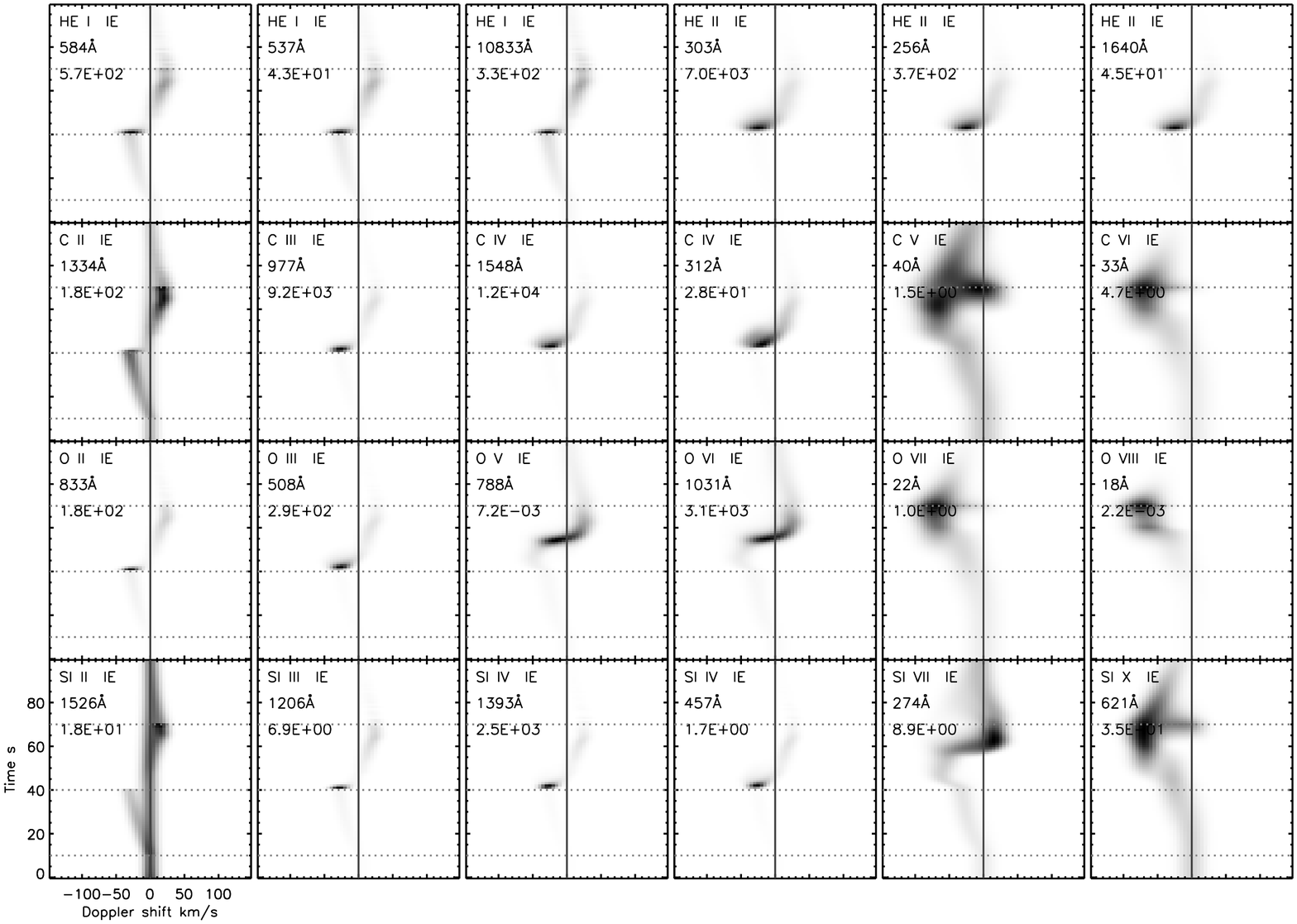}  
\caption{\label{fig:profilesae} Same as Figure~\pref{fig:profiles}, but
    for calculations assuming instantaneous ionization equilibrium. }
\end{figure}
}

\begin{document}

\title{\large The connection of Type II Spicules to the Corona}

\author{Philip G. Judge} 
\affil{High Altitude Observatory, National Center for Atmospheric Research\altaffilmark{1}, P.O. Box 3000, Boulder CO~80307-3000, USA\\ \vbox{}}

\author{Bart De Pontieu} \affil{Lockheed Martin Solar and Astrophysics Lab, 3251 Hanover St., Org. A021S, Bldg. 252, Palo Alto, CA 94304, USA}

\author{Scott W. McIntosh} \affil{High Altitude Observatory, National Center for Atmospheric Research\altaffilmark{1}, P.O. Box 3000, Boulder CO~80307-3000, USA\\ \vbox{}}
\and
\author{Kosovare Olluri}{\affil{Institute of Theoretical Astrophysics,
    University of Oslo, 
P.O. Box 1029 Blindern, N-0315 Oslo, Norway}

\altaffiltext{1}{The National Center for Atmospheric Research is sponsored by the National Science Foundation}

\begin{abstract} 
%
 We examine the hypothesis that plasma associated with ``Type II'' spicules 
  is heated to coronal temperatures, and that the upward moving
  hot plasma constitutes a significant mass supply to the solar
  corona. 
%
  1D hydrodynamical models including 
  time-dependent ionization are brought to bear on the
  problem. These calculations indicate that heating of field-aligned
  spicule flows should produce significant differential Doppler shifts between
  emission lines formed in the chromosphere, transition region, and
 corona. 
  At present, observational evidence for the computed 60-90 \velu{} differential
  shifts is weak, but the data are limited by 
 difficulties in comparing 
 the proper motion of Type-II spicules, with spectral and kinematic
 properties of  associated transition region and coronal
 emission lines. 
Future observations with the
 upcoming IRIS instrument should clarify if Doppler shifts are 
 consistent with the dynamics modeled here.  
\end{abstract}

\keywords{     Sun: atmosphere;      Sun: chromosphere;          Sun:
  transition region;  Sun: corona  }

\section{Introduction} \label{sec:introduction}

Traditionally, research on the ``coronal heating problem'' has 
focused on the transport and dissipation of non-radiative
energy directly into plasma at heights typical of the solar corona
\citep[e.g.][]{Kuperus+Ionson+Spicer1981,Parker1994,Walsh+Ireland2003},
2Mm and more above the visible solar surface.
At
a first glance, this seems odd, since the corona is a tenuous
structure fed by mass, momentum and energy from below. Indeed,
processes operating solely in coronal plasma cannot actually produce a
corona unless one already exists. Yet  five decades of detailed
spectral analysis from space experiments have implied that there is a
significant net downflow of energy in the form of heat conduction and
enthalpy from the corona towards the underlying chromosphere
\citep[e.g.][]{Mariska1992}. The evidence in support of this picture includes 
agreement between emission measures, above $10^5$ K,  from models dominated
by (downward-directed) heat fluxes, and the observation that 
the profiles  of
transition region lines, while highly variable, are nevertheless
mostly 
red-shifted.  
In addition, much magnetic (free) energy
can be stored in the coronal volume in the form of current systems
and/or waves.
Thus the community has become accustomed
to working largely within the picture of {\em in situ} dissipation of
mechanical energy directly within the coronal plasma.

However, observations from the Solar Optical Telescope
(SOT, \citealp{Tsuneta+others2008}), mounted on the stable,
seeing-free platform of the Hinode spacecraft
\citep{Kosugi+others2007} have revived some earlier ideas concerning
the coupling of the corona to the underlying chromosphere. Data
obtained and analyzed by \citet{dePontieu+others2007} have produced
quantitative information about two populations of ``spicules'', jets
of material moving supersonically above the limb and into the
corona, something that was not achievable with the previous generation
of ground-based observations \citep[e.g.][]{Roberts1945,Beckers1968,
  Beckers1972}.
De Pontieu and colleagues identified a class of finely
structured spicules with apparent motions far faster ($\sim 100$
\velu) than previously measured ($\sim 20-30$ \velu{}). These ``Type-II''
spicules, unlike their longer lived brethren, disappear in a
fashion suggesting that some of the cool material is heated during the
spicule's lifetime. Statistical correlations of these new spicules with coronal
counterparts have since been obtained using data from the Hinode,
TRACE, STEREO, and SDO spacecraft
\citep{dePontieu+others2009,McIntosh+DePontieu2009a,McIntosh+others2010,dePontieu+others2011}.

This work suggests that significant energy deposition occurs in the
plasma associated with spicular events.  Recent MHD simulations have
provided one scenario in which Lorentz forces and Joule heating on a
wide range of different field lines can explain some of the observed
phenomena \citep{Martinez-Sykora+others2011}. 
Here we test the
possibility that the heating and acceleration of plasma occurs in a 
field-aligned flow, the freshly heated spicular material 
filling the overlying atmosphere with material heated to
higher, perhaps coronal, temperatures. In other words, we test whether
a simple hydrodynamic approach can reproduce the observed
phenomena. These studies are of interest because  the
association of hot plasma with spicules suggested that
``energy deposition at coronal heights cannot be the only source of
coronal heating'' \citep{dePontieu+others2009}.

Ideas along these lines have been around for some
time. \citet{Thomas1948a}, interpreting Roberts' spicules as
hydrodynamic jets, noted that ``the directed mechanical energy of the
jet becomes converted to random thermal energy, in part, as the jet
moves though the atmosphere''. \citet{Miyamoto1949a} studied the
viscous dissipation of kinetic energy of spicular motions. But with
the observations then available, the 30 \velu{} speeds of spicules was
insufficient to produce coronal temperatures. \citet{Pneuman+Kopp1978}
estimated mass fluxes from available data, which with mass
conservation, prompted them to suggest that ``the observed [transition
region] downflow represents spicular material returning to the
chromosphere after being heated to coronal temperatures''.
\citet{Athay+Holzer1982} noted that ``spicular material is raised well
above the height that would be achieved by a projectile of the same
initial velocity, thereby obtaining gravitational potential energy
much in excess of its initial kinetic energy''.
Athay and Holzer concluded that ``if sufficient heat is added to
spicules, in conjunction with their acceleration, the spicule
phenomenon may also play a major role in the production and
maintenance of much of the solar corona."
In a paper entitled ``the coronal heating paradox'',
\citet{Aschwanden+others2007}, proposed that the the phrase ``coronal heating
problem'' is a misnomer. Instead, they argue that coronal heating
should be considered as two processes, a ``chromospheric heating
problem'' and ``coronal loop filling process'', moving the problem of
energy dissipation into chromospheric
plasma. \citet{Hansteen+others2010} present sophisticated 3D radiation
MHD calculations from sub-photosphere to corona and conclude that, for
calculations driven by the work done on photospheric fields by
convection, and dissipated by Ohmic heating, ``the heating per unit
mass [\ldots] necessarily is concentrated toward the transition region
and low corona''.


Before proceeding, we  
clarify some terminology, following
\citet{Athay+Holzer1982}. Phrases such as ``heating in the
chromosphere'', ``heating in the transition region'', as well as
``heating in the corona'' abound.  We will avoid this usage, as these
phrases seem to imply that there is something about the thermal
properties of these plasmas (e.g. ionization state, conductivities)
that delineate different mechanisms affecting the three regimes.  This
may well be the case, but few studies deal with such differences 
(one example that does is the work of \citealp{Goodman2004}).  
Also, this language suggests, perhaps subliminally, that 
energy transport 
mechanisms must be found that fit in a pre-existing thermal structure.
Instead we see the chromosphere, transition region, and
corona, including the associated zoo of observed phenomena we call
spicules, fibrils, explosive events, sprays, blinkers, etc, as {\em
  observational manifestations} of the transport and dissipation of
mass, momentum and non-radiative energy throughout the Sun's
atmosphere. When viewed in a more physical sense, one
avoids difficulties arising from observationally-based language use,
such has, how can so much heating appear to be confined to the thin
solar transition region? How can an observational phenomenon like a
spicule can be considered as a source of mechanical heating? Why need
one consider coronal heating to be paradoxical at all?

\section{Calculations}

The work of \citet{dePontieu+others2009, dePontieu+others2011} suggests that significant
energy deposition occurs in plasma associated with
spicules.  Nearly all models on spicules 
``begin with some form of deposition of energy in the photospheric or
chromospheric portion of a magnetic flux tube which extends from the
photosphere into the corona'' \citep{Sterling2000}, a notable exception being
the recent work of  \citet{Martinez-Sykora+others2011}.   
In
their detailed calculations, the upward jet motion is essentially a
pressure-gradient-driven field-aligned flow,  initiated by the horizontal compression of 
plasma by the Lorentz force associated with emerging flux 
causing strong Joule heating in and around the
jet-like feature. Here, we will study 
the response of the atmosphere to {\em ad-hoc} forcing and 
heating terms.  Since the Type II
spicules are very thin (high aspect ratio),  as a first attempt to
compare simulated line profiles with observations, 
we perform 1D gas dynamic calculations
solving the set of physical equations described by
\citet{Hansteen1993}.  These equations follow the time evolution of
a single fluid's mass, momentum, total energy, and also the number
densities of the ground levels of atoms and ions of 
H, He, C, N, O, and Si. The adopted ionization
and recombination rate coefficients, and the special treatment of
ionization of hydrogen, are described in the Appendix.  

Sources
and sinks for momentum and energy include gravity and explicit heating
per proton and neutral hydrogen atom.  Radiative losses in the energy
equation were treated  using the computed time-dependent population
densities for 
H, He, C, N, O, Si, losses from other elements and
free-free emission were treated using a lookup table computed by PGJ using the DIPER package 
\citep{Judge2007a}. All these calculations are  based upon bound-bound collision
strengths mostly from the CHIANTI project \citep{Landi+others2006}. 
Heat conduction is included using the Spitzer
formula. 
The internal energy of the gas is computed using the translational and
internal degrees of freedom, i.e. the usual perfect gas energy density
plus the ionization energy.    The internal energy stored only in H and He
ionization is included.  For H, the internal energy is modified by the
Balmer continuum radiation and so depends on radiative transfer.
In the
chromosphere we set it to be just $\frac{3}{4}$ of the ionization energy, as 
$\frac{1}{4}$ of this energy is supplied by photoionization from the
$n=2$ levels by
photospheric radiation. 
This correction has a minor effect on the calculations.  

The equations, which we solve in their conservative forms, 
are of mixed character. Advection dominates the time
evolution of all equations except when the plasma reaches coronal
temperatures, where the non-linear diffusive heat conduction term
dominates the energy flux.  We therefore adopted an operator splitting
time integration scheme \citep[e.g.][]{Hansteen+others2010},
integrating the advection terms using a Lax-Friedrich finite
difference scheme,
accurate to first order in
space and time. 
(We tried slope-limited higher order schemes which are
more accurate for smooth solutions. But these 
introduce interpolation errors when, as in the
transition region modeled here, gradients are very steep).  
After each advection step, 
we apply a Crank-Nicolson scheme to integrate up the
temperature changes due to conduction.
A uniform, fixed grid of 4097 points was used ($\equiv 10$ km
resolution) along a semi-circular loop of total length 20 Mm (footpoints
separated by 40/$\pi$ Mm) with the dense ``photosphere" at both
footpoints.  Boundary values at the photosphere were held fixed.

\subsection{Initial state}

An initial state was constructed in which protons and hydrogen atoms
were heated at the rates
\begin{eqnarray}
  \label{eq:1}
  \epsilon_p &=& 2\times10^{-12} n_p  \ \ \ {\rm erg~cm^{-3}~s^{-1} } \\
\epsilon_H &=& 1\times10^{-14} n_H  \ \ \ {\rm erg~cm^{-3}~s^{-1} }
\end{eqnarray}
where $n_p$ and $n_H$ are number densities of protons and neutral
hydrogen atoms respectively.  These values produce a chromosphere and
a corona with a peak temperature of $10^6$K.  
The equations for conservation of mass, momentum and
energy were evolved, starting from model C from
\citet{Vernazza+Avrett+Loeser1981}, together with rate equations for
the ground states of neutral H and and He, He$^+$, and for H$^+$ and 
He$^{2+}$.  
Figure~\pref{fig:eqm} shows the configuration of the starting
solution. It is close to
equilibrium, but has a small yet persistent subsonic downflow in the lower transition region.  

\subsection{Forcing}

Guided by the observational analyses of 
\citet{dePontieu+others2009, dePontieu+others2011}, 
the initial state was forced in an {\em ad-hoc} way 
by accelerating chromospheric plasma 
and heating it for $t_1$  seconds, 
followed by another $t_2$ second period of
more intense heating but no acceleration (coasting).  
Here we discuss two calculations, with parameters listed in table 
\pref{tab:parameters}.   In Calculation ``A'', plasma initially
between 2000 and 2400 km in height is heated and accelerated for 30
seconds, and then simply heated for another 30 seconds.  Calculation B has
shorter
acceleration and heating phases (15 second durations), heats/accelerates plasma 100 km lower
and higher (roughly one chromospheric scale height), but it is heated in the
coast phase by a larger factor.

The heated plasma was
initially located at the top of the chromosphere, extending into the
low corona, and as the calculations
evolved the same plasma was heated by tracking the plug of plasma.
The net upward acceleration of the plug was set to 
2 km~s$^{-2}$, about four times the solar gravitational acceleration.  
The heating rates for both neutrals
and protons were increased three-fold during the first period, to
counteract cooling by adiabatic expansion, and then by much larger factors
during the second period.  In this way we mimic the
heating of plasma already accelerated to $\sim35$ and $\sim25$ \velu{} to coronal
temperatures for A and B respectively, a little less than speeds of
typical spicule observations of 
\citet{dePontieu+others2009,McIntosh+DePontieu2009a}.  In order to avoid unphysical
discontinuities, the acceleration and heating were smoothed using a Gaussian spatial
profile of width (FWHM) 330 km about the initial state.

Figures \pref{fig:end} and \pref{fig:afterend} show the atmospheric
parameters 20 seconds after the beginning of the acceleration, and
then 20 seconds after  the strong heating began, for calculation A. 
The sound crossing time for the calculation shown is about 150s.  Note
that, unlike the far more sophisticated calculation of 
\citet{Martinez-Sykora+others2011}, these two
calculations do not produce long, cool spicule-like structures, a well
known problem with most spicule models \citep{Sterling2000}.  

Our parameter choices parameters are arbitrary in physical terms.  But
they produce results typical of those solutions in which chromospheric
plasma is accelerated and heated in the fashion broadly suggested
by \citet{dePontieu+others2009, dePontieu+others2011}.  Forcing the 
calculations in
denser, deeper layers produces warmer chromospheres with no fast
outflow
to the corona; forcing them higher produces insignifcant chromospheric dynamical
effects.

\subsection{Results}
First we discuss calculation A. 
The response of the atmosphere to the input of momentum and energy is
simple: the plug of forced plasma initially expands mostly outwards
into the tenuous coronal plasma reaching speeds of $\sim35$\velu.  After this first phase the additional extreme heating
provides, through the associated over-pressure, additional outward
acceleration of coronal plasma to near 90\velu.    In the uppermost
regions of the chromosphere, some downflow is seen as the high pressure
plug pushes downwards, as discussed in the MHD calculations of 
\citet{Hansteen+others2010}.
Results for calculation B are similar, except that the extra heating 
leads to upflow speeds of $\sim 120$ \velu{}. Also, a cool ($\sim
3\times10^4$K), dense plug
of
plasma is driven into the corona in calculation B behind the 
shock wave propagating upwards into the corona.

We have investigated the influence of certain assumptions in our calculations.
Our results are sensitive to the treatment of radiation losses.  We
treat the losses as effectively thin, as in most earlier work, but there
 are two significant, physically distinct effects that preclude
our use of standard radiation loss curves
\citep[e.g.][]{Raymond+Smith1977}.  The first is to account for
photoionization of hydrogen by photospheric Balmer continuum
radiation.  Non-LTE radiative transport is beyond the scope of the
present paper \citep[cf.][]{Carlsson+Stein1995}, but the qualitative
effects of Balmer continuum photoionization on radiation losses have
been studied by \citet{Athay1986}.  Ionization via Balmer continuum
photoionization lowers the radiation losses from hydrogen L$\alpha$,
a line that otherwise can dominate the plasma radiation losses between
$10^4$ and $2\times10^4$K.  
 The second effect is that for the ions
treated in detail, we solve radiation losses without assuming
ionization equilibrium.  Our treatment of hydrogen is discussed in the
Appendix.  The two effects both serve to increase the ionization of
hydrogen relative to radiation-free optically thin conditions,
allowing more energy to propagate upwards  into the low corona
that otherwise would
have been radiated by the chromosphere.

\subsection{Line profiles}

To compute profiles of selected EUV emission lines, we take the 
time-dependent 
populations of all ions of H, He, and of all charged ions of C, N, O, and
Si$^{1+}$ through  Si$^{8+}$, and use these populations to compute the
emitted power in selected lines, which are then
integrated along a given line of sight 
to compute the line profiles assuming optically thin
conditions.   The same data are used in the calculation of radiative
cooling for these ions.
The approach of solving in time just for the ionization states of the
ions, ignoring excited levels (except for the special case of hydrogen), 
is justified because 
for the ions of interest, the ion densities are accurately tracked in
time and space and the emitted radiation can be computed  {\em
  post-facto} because of the separation of timescales involved
\citep{Judge2005}.   We compute profiles using Gaussian profiles
including both thermal and a 10 \velu{} turbulent component to the
line widths, and include the Doppler shifts of the moving plasma as seen
vertically.    To make meaningful comparisons with observations we average
all the emission which, when observed vertically, lies within 2.175Mm
(3\arcsec{} as seen from earth) 
of the loop footpoint projected on the solar surface.   This length
was chosen simply because the ``transition region'' moves up and down 
significantly in our calculations \citep[cf.][]{Hansteen1993}, and
choosing a smaller length scales can produce artifacts as the
transition region emission moves in and out of this range.  (Also, the
point spread functions  of 
the best EUV spectrometers would have contributions wider than $1\arcsec$).
Figure \pref{fig:profiles} shows the resulting line intensity profiles
of a variety of lines of He, C, O and Si as a function of Doppler shift
and time, for calculation A, in which additional heating
and acceleration starts at $t=10$ sec. In each panel, the element, central wavelength and
maximum  (wavelength-integrated) intensities are listed.  The figure shows
profiles of lines with increasing atomic number, from top to bottom, and
with increasing ionization stage- and hence typical formation
temperature- from left to right.    The intensities
calculated
generally exceed those measured in extant low-resolution
observations, but given that the filling factor of type II spicules 
may  be very  
small \citep[e.g.,][]{dePontieu+others2007}, we do not consider this an
important difference.  

The overall dynamics is
reflected in the lines of Si, from the chromospheric Si II 1526 \AA{}
line, through the Si X 621 \AA{} coronal line.  The upward
acceleration of chromospheric plasma between 10 and 40s can be clearly
traced in the calculations for the Si II line, and in the 
C II line.  It is also visible in the coronal lines (C V, C IV, O VII,
O VIII, Si X).  But in these lines 
additional
acceleration due to the pressure gradient in the hotter plasma
extends for an additional 30 sec or so.

In fact the chromospheric and
coronal lines behave quite differently: the blueshifts of the
coronal lines are {\em systematically larger} than chromospheric and
  transition region lines by some 50-80 \velu{} . The line profile of
Si II for example shows chromospheric  acceleration, but the
blue-shifts disappear with the onset of the additional heating at
$t=40s$.    In contrast, the coronal lines are 
blueshifted relative to the Si II line when $t> 40s$.
After the forcing period the acoustic wave generated  propagates away from the
footpoint, and all the lines become dimmer.  In addition, all
chromospheric lines return near $t=100$s to zero Doppler shift and even
become red-shifted (faintly seen in the lines of helium, for
example). 

The different Doppler shifts calculated for chromospheric and coronal
lines, found in this particular calculation, is a more general result.
It is seen  in calculation B, where the stronger heating
rates produce sudden pressure-driven acceleration, so that the coronal
line blueshifts exceed the Doppler shifts of chromospheric lines by
$\sim100$ \velu{} (Figure \pref{fig:profiles1}).  Several other calculations
with different heating rates higher and lower in the chromosphere
produce qualitatively similar  results.

\subsection{Other spectral signatures of the dynamics of heated outflowing plasma}

Another systematic difference is seen in the emergent spectra we 
have calculated.  
The spectral lines we selected include some atomic transitions which have
sensitivity to ionization non-equilibrium effects, through their
relatively high excitation energy.  Excitation of helium atoms and
ions requires a large energy (compared with typical thermal energies
found under ionization equilibrium conditions) because the excited
levels have a larger principal quantum number $n$ than the ground
levels.  Any excitation requires a ``$\Delta n \ge 1$'' transition.
In contrast, other atomic ions can have both $\Delta n=0$ and $\Delta
n \ge 1$ transitions.  The plotted 312 and 457 \AA{} lines of C IV and
Si IV respectively are $\Delta n=1$ transitions ($2s-3p$ and $3s-4p$),
unlike the 1548 and 1393 \AA{}  lines which are $2s-2p$ and $3s-3p$
transitions respectively.

In Figure \pref{fig:profilesae} we show the profiles computed with the 
full dynamics, but assuming instantaneous ionization equilibrium 
(IE).  Comparison with 
Figure \pref{fig:profiles} clearly reveals
departures from IE. The IE calculations are usually more variable as it takes some time
for ions to become ionized or to recombine in response to changes in
temperature and density.   But, systematically, 
the $\Delta n=1$ transitions are {\em all 
brighter} than IE would predict.  This effect
results from plasma becoming ionized, under conditions where 
excitation and radiative decay of the levels occurs before the ion to which they belong
becomes significantly ionized.  In helium lines this has been invoked
to explain their anomalous brightness  
\citep[e.g.][]{Jordan1975,Laming+Feldman1992,Pietarila+Judge2004,Judge+Pietarila2004,Judge2005}.   
A similar effect is seen in the C IV and Si IV ions, but here we can
compare directly the intensities $\Delta n=1$ with the $\Delta n=0$
transitions.  In both cases, when the emitting 
plasma is undergoing heating, the $\Delta n=1$ transitions become
enhanced relative to the $\Delta n=0$ transitions by a factor of 3.1 and
2.5 for Li-like C IV and Na-like Si IV respectively.  

Such large departures from ionization equilibrium should be measurable.
Simultaneous observations of the $\Delta n=0,1$ transitions in C IV
and Si IV have presumably been obtained using the CDS and SUMER
instruments on the SOHO spacecraft, but it is beyond the scope of the
present paper to identify and analyze such data.


\section{Discussion}

Our 1D
hydrodynamic calculations attempt to capture the evolution of the
emitted spectrum of a plug of accelerated and heated plasma flowing
along magnetic field lines extending upwards from the
upper chromosphere.  The dynamics and emergent 
spectra are  computed using full
non-equilibrium in the atomic rate equations.
The calculations reveal that the coronal emission
which accompanies the heating needed to produce coronal plasma from
the cool spicular plasma
is blueshifted by $\sim 50-90$ \velu{} relative to the chromospheric
emission (compare lines of 
He~I, C II Si II with the coronal lines in figs.~\pref{fig:profiles} \pref{fig:profiles1}). 
This is because, in the hydrodynamic regime, the heating produces an over-pressure 
and a subsequent expansion of the plasma into the corona.  Thus, if
indeed the chromospheric plasma in Type II spicules is injected into
the corona, at the same time being heated in this fashion, then we must expect the
velocity distributions of chromospheric and coronal lines to differ.
Accompanying the prediction of Doppler shift differences, 
changes in the intensities of
line ratios that are traditionally viewed as being sensitive to
electron temperature, are also predicted in our rapidly heated outflow models.
These intensity changes result from the slowness of 
ionization/recombination processes relative to dynamical timescales.  
It would seem worthwhile to 
investigate atomic systems of the Li and Na-like isoelectronic 
sequences observationally, following earlier work \citep{Heroux+others1972}
as well as other atomic systems with temperature  sensitive line ratios
\citep[e.g.][]{Pinfield+others1999}.

%
%

Such a systematic difference between cool and hot flows associated
with Type II spicules 
has not been reported in
the observational literature, but we note that the
observations have significant uncertainties.  
The cool components of Type II spicules at the limb show apparent
velocities of order 50-100 \velu{} \citep{dePontieu+others2007}, with
rapid blueshifted events, thought be their disk counterparts, showing
line-of-sight velocities of order 50 \velu{} \citep{Rouppe+others2009} and
apparent velocities of order 75 \velu{}
\citep{dePontieu+others2011}. 
The velocities of coronal counterparts are also difficult to
determine precisely. They are derived from 
weak blueshifted emission
components, seen at the base of much stronger emission cores, near 
the limit at which EUV spectrographs can operate (e.g., instrumental
broadening, signal-to-noise, ...). These measurements suggest line-of-sight
velocities of order 50-150 \velu{} for the coronal
counterparts of spicules. 
Because both the chromospheric and coronal measurements are
significantly impacted by viewing geometry, radiative transfer and
instrumental limitations, 
it is not yet clear that the observed velocity distributions are
sufficiently well determined to reveal a velocity difference of 50 \velu{} between
the chromosphere and coronal counterparts of spicules.

If the large relative shift in the Doppler motions predicted 
here are not confirmed by future high resolution measurements (such as
those from the IRIS mission\footnote{http://iris.lmsal.com/index.htm}), 
we would conclude that the suggestion put forward by
\citet{dePontieu+others2007,dePontieu+others2009,dePontieu+others2011},
namely the close correspondence of velocities from spicules with those
of Doppler shifted lines in the transition region and corona, 
cannot credibly be accounted for in the type of 1D hydrodynamic flow modeled
here.   
One possible resolution might simply be that the
magnetic fields within the tube are significantly 
twisted such that the spicular plasma has an azimuthal as well as axial component as the
Lorentz force directs the plasma along field lines.  In this case the
Doppler shifts computed should be multiplied by the cosine of the
magnetic pitch angle.  

The recent MHD calculations of \citet{Martinez-Sykora+others2011}
might help resolve the problems identified here, in that the system 
modeled is far more complex than our simple
calculations and does not seem to show such large velocity differences
between chromospheric and coronal lines.  In their work, they find that chromospheric
material is injected into the corona via, first, compression due to
horizontal Lorentz forces associated with emerging flux, and second,
via the field-aligned gas pressure gradient.  As the field-aligned
flow progresses it is heated via Joule dissipation, in a fashion 
not dissimilar from the
ad-hoc calculations we present.  However, in the modeled MHD system,
the acceleration and heating acts, with varying strength, on plasma
that occurs on an ensemble of neighboring field
lines. As a result, the coronal emission 
has contributions from many magnetic field lines with varying mixes of
chromospheric and coronal plasma/flows.  Thus, the line profiles of coronal plasma must be
carefully synthesized to find the relative contributions of
these different components.  Notably though, the cool plasma that actually
flows into
the corona in these MHD calculations is indeed a field-aligned
flow.  Thus, 
if our parametrized heating and acceleration profiles are
representative of what occurs in the complex 3D configuration,
the heated plasma in that part of the calculated flow 
would be expected to behave like the
calculations presented in the present paper. Current EUV observations
lack the spatio-temporal resolution to resolve the finely structured
dynamics and energetics predicted by the model of
\citet{Martinez-Sykora+others2011}.


The potential problem of Doppler shift differences may
be avoided in an entirely different scenario \citep{Judge+Tritschler+Low2011},   where
it was suggested that 
Type II
spicules, in particular those 
with apparent upward velocities in excess of 50 \velu{},
correspond not to jets but to the line-of-sight superposition of 
warped current sheets modulated by 
Alfv\'enic\footnote{Simply meaning motions dominated by the balance between
  magnetic tension and inertial terms.}
 fluctuations driven from below. Low-frequency Alfv\'enic-type wave motions
have been reported before in movies of Type II spicules
\citep[][]{DePontieu+others2007b}. 
  Should Type II spicules be proven to be sheets, the
calculations here are largely irrelevant, and the physical 
connection between the chromospheric, transition region and coronal
emission proposed earlier would be called into question.  
But, as suggested by \citet{Judge+Tritschler+Low2011}
the 1D flow
picture might  apply to a subset of observed spicules with plasma 
genuinely flowing at lower speeds. 
However, several questions regarding the overall interpretation of 
spicule properties at the limb, their relation to RBEs,  and
statistical analyses would then arise.


Whatever the outcome, 
we conclude that the role of Type II spicules in supplying mass and
 energy to the corona
 \citep{dePontieu+others2007,dePontieu+others2009,dePontieu+others2011},
is a subject ripe for future study.  
Further observational work on spicule properties and their relationship to 
the properties of the associated hotter plasma seems warranted.  
The upcoming IRIS mission
 seems well positioned to address this question.

\acknowledgments We are grateful to Paul Cally and an anonymous
referee
for a careful reading of the
paper and for helpful comments.

\appendix

\section{Atomic rate equations}

The system of equations integrated in time includes equations for the
number densities of atoms and atomic ions.  In 1D, these take the form 
\begin{equation} \label{eqn:rate_equations}
\ddt{\pop_i\frt} = \sum^N_{j \ne i}\pop_j\frt{}P_{ji}\frt{} 
- \pop_i\frt{} \sum^N_{j\ne i} P_{ij}\frt{}.
\end{equation}
where $t,x$ are time and distance respectively, $\pop_i$ is the
population density of atomic state $i$, and the coefficients $P_{ji}$ represent the
transition rate, units s$^{-1}$, from state $j$ to state $i$ (and vice
versa).     Only long-lived states need tracking in time
\citep{Judge2005}, so that the only states we solve for are the ground
states of atomic ions and bare atomic nuclei.     Given the solar 
radiation field and local thermal properties in the corona, the
largest contributions to coefficients $P$
are ionization and recombination coefficients, by electron impact 
(charge transfer collisions with H and He are sometimes important 
but are not included here).  Here we adopt the ionization rate
coefficients of \citet{Arnaud+Rothenflug1985}, and fits to
recombination rates in the form tabulated by
\citet{Shull+vanSteenberg1982}, computed by one of us (PGJ) from the
detailed photoionization cross sections computed by the OPACITY project
\citep{Seaton1987}.   These differ substantially from, and should be
more accurate than, the calculations
of \citep{Shull+vanSteenberg1982}.  A comparison of recombination 
rates computed from the OPACITY project and more detailed work 
of \citet{Nahar+Pradhan1992} is given in
\citep[][Fig. 13]{Judge2007a}\footnote{Available at 
http://www.hao.ucar.edu/modeling/haos-diper/.}.  The systematic differences show the 
rates to agree within 0.2dex with some excursions of a factor of 2
possible.  

Hydrogen ionization occurs not quite so simply as described 
above, and it is treated differently.  This is because the $n=2$
levels can have a significant population owing to scattering in
Ly$\alpha$
within the chromosphere, and Balmer continuum radiation from
the upper photosphere can photoionize hydrogen
\citep[e.g.]{Hartmann+MacGregor1980, Vernazza+Avrett+Loeser1981}. Thus we include 
photoionization from the $n=2$ levels. We compute the $n=2$
populations 
assuming that Ly$\alpha$ is
in detailed balance in the chromosphere (mostly neutral) but that the
population is greatly reduced below this limit in the corona.  As an {\em ad-hoc},
qualitative representation of these effects we apply the non-LTE factor 
$\left ({ n_1 \over {n_1+n_\kappa}} \right)^2$:
\begin{equation}
  \label{eq:fudgeh}
 { n_2 \over n_1} = {n_2^* \over n_1^*} \left ({ n_1 \over {n_1+n_\kappa}} \right)^2
\end{equation}
where the asterisk refers to LTE populations, and $n_\kappa$ is the 
population density of protons.  We adopt a
photoionization rate of $8\times10^3$ s$^{-1}$ from the $n=2$ level
\citep{Vernazza+Avrett+Loeser1981}.  We assume radiative detailed balance
throughout in the Lyman continuum and thus set photoionization and
recombination rates to
zero in that transition.   This is not a serious problem in that it is
correct in the chromosphere, and photoionization is not very important  
throughout the transition region and corona, relative to electron
impact ionization.

\tabone

\figone
\figtwo
\figthree
\figfour
\figfive
\figsix
\end{document}